
%
%
%
%
\magnification=1200
\null
\hsize=5.9truein
\vsize=8.5truein
\baselineskip=14pt
\nopagenumbers
\overfullrule=0pt

\def\gs{{g^{\prime\prime}}}
\centerline{{\bf{LIMITS ON THE BESS MODEL AT NLC}}
\footnote*{Invited talk at the ``Workshop on Physics and Experiments at 
Linear $e^+e^-$ Colliders'', Waikoloa, Hawaii, April 26-30, 1993 (presented 
by D. Dominici)}}
\vskip 1.5truecm
\baselineskip=10pt
\centerline {R. CASALBUONI, S. DE CURTIS, D. DOMINICI}
\centerline{\it {Dipartimento di Fisica, Univ. di Firenze, Firenze Italy}}
\centerline{\it{ I.N.F.N., Sezione di Firenze, I-50125 Firenze, Italy}}
\medskip
\centerline {P. CHIAPPETTA}
\centerline{\it{ CPT, CNRS, Luminy Case 907, F-13288, Marseille, France}}
\medskip
\centerline {A. DEANDREA, R. GATTO}
\centerline{\it{ D\`ept. de Phys. Th\`eor., Univ. de Gen\`eve, CH-1211 
Gen\`eva 4}}
\vskip 0.5truecm
 
\baselineskip=14pt
\vskip 1.0true cm
\newcount \nfor

\def \form {\global \advance \nfor by 1 \eqno(1.\the\nfor)}
\noindent
 
{\bf 1. The BESS model and the present bounds on its parameters}
\vskip .5truecm
 
This contribution to the workshop contains an update of our
previous work$^{1,2}$ concerning the sensitivity of future $e^+e^-$
linear colliders to a strongly interacting electroweak
sector. We have examined the occurrence of new effects
at proposed colliders, ranging from an energy
of $300~GeV$ to $2~TeV$ and of different luminosity options.
 
Our calculations are performed within a 
strongly interacting electroweak symmetry breaking
model, named BESS$^3$. The model is an effective lagrangian 
parameterization of the symmetry breaking mechanism,
based on a symmetry $SU(2)_L\times SU(2)_R/SU(2)_{L+R}$. New vector
bosons are present and we expect new
effects due to their mixing
with the electroweak gauge bosons and their fermion
couplings.
 
The parameters of the BESS model
are  the mass of
these new bosons $M_V$, their self coupling
$\gs$  and a third parameter $b$ whose strength characterizes
the direct couplings of the new vectors $V$ to the 
fermions. However due to the mixing of the $V$ bosons 
with $W$ and $Z$, the new particles are coupled to the
fermions also when $b=0$. The parameter $\gs$ is expected
to be large due to the fact that these new gauge bosons
are thought  as bound states from a strongly interacting electroweak
sector.   By taking the formal $b\rightarrow 0$ and
$\gs\rightarrow\infty$ limit, the new bosons decouple 
and the standard model (SM) is recovered. Whereas by considering only
the limit $M_V\rightarrow\infty$ they do not decouple.
 
In our study we have considered the
sensitivity of processes involving
different final states from $e^+e^-$ to the new physics
predicted by the BESS model.
 
Future $e^+e^-$ colliders can be very useful in studying 
detailed  properties of the vector particles predicted
by the BESS model or by extra $Z^\prime$ models, particularly after
their possible discovery at future $pp$ colliders like LHC or SSC.
If the energy of the collider is very close to the mass of the resonances, then 
they can be copiously produced and studied. Also note that, in such linear
colliders, beamstrahlung will automatically provide for a spectrum of
 lower initial energies.
 
In  our analysis we consider
mainly indirect effects of the new particles 
to different observables, supposing $M_V>\sqrt s$.
We have considered $e^+e^-$
annihilation into fermion
antifermion and two gauge bosons.
In this processes the main effect is the presence
of the new neutral vector exchange in the 
$s$ channel.  
We have also considered the fusion channel where
two gauge bosons and two leptons
appear in the final state.  In this process both charged and neutral 
vector bosons can be produced in the $s$ channel fusion.
 
These seem to be the most promising channels, expecially
the annihilation, where all the initial state energy
is converted in the energy of the final state. The $WW$ annihilation 
channel is particularly
relevant, especially if the $W$ polarization can
be reconstructed, because  $V$ particles 
are strongly coupled to longitudinally polarized $W$'s.
The $W_LW_L$ channel severely constrains the parameter
space of the model, already for a $500~GeV$ collider.
The bounds improve by increasing the energy, because
the deviations due to BESS increase with the energy.
 
The fermion channel is relevant only for
energies in the range $300-500~GeV$ and by using
longitudinally polarized electron beams.
 
The fusion channel can be interesting for higher energies
$1.5-2~TeV$, however an extremely high
luminosity is needed.
 
We have not considered the $\gamma\gamma$ and the $e\gamma$
options, because  the new gauge bosons are exchanged in the $t$
channel and the couplings of the $W$ to a photon as well as 
the fourlinear $WW\gamma\gamma$ couplings are the same as
in the SM.
 
Finally, we have not taken
into account beamstrahlung effects. However for two body final states, 
which are the most successful channels in our analysis,
the practical effect is a reduction of the luminosity. 
This means that, with the assumed nominal luminosity, one has to run for a
correspondingly longer period.
 
 Using the following LEP data averaged on the four LEP experiments and the
CDF/UA2 measurement of the  mass ratio $M_W/M_Z$ $^4$
$$
\eqalign{
&M_Z=91.187\pm 0.007~GeV\cr
&\Gamma_Z=2488\pm 7~MeV\cr}
$$
$$
\eqalign{
&\Gamma_h=1740\pm 6~MeV\cr
&\Gamma_{\ell}=83.52\pm 0.28~MeV\cr
&A_{FB}^{\ell}=0.0164\pm 0.0021\cr
&A_{\tau}^{pol}=0.142\pm 0.017\cr
&A_{FB}^{b}=0.098\pm 0.009\cr
&{M_W\over M_Z}=0.8798\pm 0.0028}
\form
$$
we can derive 
bounds on the BESS model, that we express as $90\%$ C.L.
contours in the plane $(b,g/\gs)$ for given $M_V$ (see Fig. 1).
The predictions of these observables for BESS are evaluated by including
the same one-loop electroweak radiative corrections of the SM calculated
with $M_H$ interpreted as a cut-off $\Lambda$.
The bounds depend
mainly on the still large  range allowed for the SM parameters
$m_{top}$ and $\alpha_s$. For this reason in Fig. 1 we show 
the total allowed region for $m_{top}$ and $\alpha_s$ within the indicated 
ranges.
 
 
\baselineskip=10pt
\smallskip
\noindent
{\bf Fig. 1} - {\it  $90\%$ C.L. contour in the plane $(b,g/\gs)$
for $M_V=600~GeV$, from LEP and CDF/UA2 data
$(130\le m_{top}(GeV)\le 180$, $0.11\le\alpha_s\le 0.13$, and $\Lambda=1~TeV)$.}
\smallskip
 
\baselineskip=14pt
 
The bounds become stronger for increasing 
$\alpha_s$ and  $m_{top}$, while they are almost independent 
of the mass of the new
resonances $V$ and of the choice of the cut-off. 
 
LEP 200 is expected to increase only marginally the sensitivity over LEP.
The relevant modification will be brought by the more accurate determination
of $M_W$.       
\vfill\eject
\newcount \nfor

\def \form {\global \advance \nfor by 1 \eqno(2.\the\nfor)}
\noindent
{\bf  2. Annihilation channels}
\vskip .5truecm
 
Our analysis, in the fermion channels, is 
based on the following observables: the total hadronic 
and muonic cross sections, 
their ratio, 
the forward-backward asymmetries in muons and $\bar b b$
 and, assuming a longitudinal polarization $P_e$ of the electron beam, the 
left-right asymmetries in muons, $\bar b b$ and hadrons
$$
\eqalign{
&\sigma^{\mu},~~\sigma^h,~~R=\sigma^h/\sigma^{\mu}\cr
&A_{FB}^{e^+e^- \to \mu^+ \mu^-},~~ A_{FB}^{e^+e^- \to {\bar b} b}\cr
&A_{LR}^{e^+e^- \to \mu^+  \mu^-},~~A_{LR}^{e^+e^- \to {\bar b} b},~~
A_{LR}^{e^+e^- \to {had}} \cr
}\form
$$
We assume a systematic error in luminosity 
${{\delta {\cal L}}/ {\cal L}}=1\%$ and ${{\delta\epsilon_{hadr}}/
\epsilon_{hadr}}=1\%$ (which is perhaps an optimistic choice due to the 
problems connected with the $b$-jet reconstruction),
$\delta\epsilon_{\mu}/\epsilon_{\mu}=0.5\%$,
where $\epsilon$ denote the selection efficiencies. 
 
Concerning the $WW$ channel,  we study the following observables:
$$
{d\sigma \over {d\cos\theta}}(e^+ e^-\to W^+ W^-),~~~~~~
 A_{LR}^{{ e^+ e^- \to W^+ W^-}}
\form
$$
where $\theta$ is the center of mass scattering angle. We have also
considered the possibility to measure the final $W$ polarization
by using the
$W$ decay distributions, and we have added
to our observables the longitudinal and transverse polarized $W$
differential cross sections and asymmetries.
 
In order to get a clear signal to reconstruct the polarization of the $W$'s
we study the channel for one $W$  decaying leptonically 
and the other hadronically. We have assumed 
an effective branching ratio $B=0.1$ to take into account 
the loss of luminosity from 
beamstrahlung$^5$.
 
The analysis has
been performed by taking 19 bins in the angular region restricted by
$|\cos\theta|< 0.95$ and assuming systematic errors 
${{\delta B}/ B}=0.005$$^{1,2}$
 and $1\%$ for the acceptance.
 
The contours shown in Fig. 2 correspond to the regions which are allowed
at 90\% C.L. in the plane $(b,g/\gs)$, assuming that the BESS deviations
for the considered observables from the SM predictions are within the 
experimental errors. Here we assume $\sqrt{s}=300~GeV$ with an integrated
luminosity of $L=20~fb^{-1}$ and $M_V=600~GeV$
(all the following results are obtained for $m_{top}=150~GeV$ 
and $\Lambda=1~TeV$).
 
By comparing with the present bounds given in Fig. 1 we see that the 
combination of all the observables provides for an efficient test at a collider
with $\sqrt{s}=300~GeV$.
 
 
\baselineskip=10pt
\smallskip
\noindent
{\bf Fig. 2} - {\it  $90\%$ C.L. contours in the plane $(b,g/\gs)$
for $\sqrt{s}=300~GeV$ and $M_V=600~GeV$
 from the unpolarized
       $WW$ differential cross section (solid line), from  the 
       $W_{L}W_{L}$ differential cross section
       (dashed line), from all the 
       differential cross sections for $W_{L}W_{L}$, $W_{T}W_{L}$, 
       $W_{T}W_{T}$ combined with the $WW$ left-right 
       asymmetries (dotted line) and from all the WW and fermion
       observables with $P_e=0.5$ (dash-dotted line). 
	The allowed regions are the internal ones.}
\smallskip
 
\baselineskip=14pt

 
\baselineskip=10pt
\smallskip
\noindent
{\bf Fig. 3} - {\it  $90\%$ C.L. contours in the plane $(M_V,g/\gs)$ for 
	 $\sqrt s=0.3,~0.5,~1~TeV$, $L=20~fb^{-1}$ and $b=0$.
       The solid line corresponds to the bound from
       the unpolarized $WW$ differential cross section, 
       the dashed line to the bound  
       from  all the polarized
	 differential cross sections $W_{L}W_{L}$, $W_{T}W_{L}$,
      $W_{T}W_{T}$   combined with 
       the $WW$ left-right asymmetries.
       The lines give the upper bounds on $g/\gs$.}
 
\smallskip
 
 
\baselineskip=10pt
\smallskip
\noindent
{\bf Fig. 4} - {\it  $90\%$ C.L. contours in the plane $(\sqrt{s},g/\gs)$ for 
	 $M_V=1.5~TeV$, $b=0$ and
       $L=20~fb^{-1}$. The lines corresponds to the same choice of 
       observables of Fig. 2 and represent the upper bounds on $g/\gs$.
       The black dots are the bounds for 
       the unpolarized $WW$ differential cross section and from
       all the WW and fermion observables by considering 
       $\sqrt{s}=1~TeV$ and $L=80~fb^{-1}$.}
 
\smallskip
 
\baselineskip=14pt
 
By increasing the energy of the collider, the allowed region of
Fig. 2 shrinks$^2$
and the fermions observables become irrelevant. 
This can be seen in Figs. 3, 4. It is due to the fact that in BESS
the vector resonances are strongly coupled to the longitudinal vector
bosons and this interaction destroys the cancellation among the $\gamma-
Z$ exchange and the neutrino contribution occurring in the SM.
 From these figures it also appears the  relevance of final $W$'s 
polarization reconstruction. 
 
\vskip 1truecm
\newcount \nfor

\def \form {\global \advance \nfor by 1 \eqno(3.\the\nfor)}
\noindent
{\bf  3. Fusion subprocesses}
\vskip .5truecm
 
The fusion subprocesses are interesting as they allow 
for study of a wide range of mass for the $V$ resonance for a 
given $e^+e^-$ c.m. energy. 
 
We have studied the
production of  $W^+ W^- $ pairs by the fusion mechanism of a pair
of ordinary gauge bosons, each being initially emitted from an electron or a 
positron.
 In the so called effective-W approximation the initial $W,Z,\gamma$
are assumed to be real and the cross section for producing a  $W^+ W^- $ pair
is obtained by a convolution of the fusion subprocess with the luminosities of
the initial $W,Z,\gamma$ inside the electrons and positrons. 
 
We expect big deviations in the invariant mass differential
cross section \break
$d \sigma/d M_{WW}$ in all the processes mediated by 
the exchange of the 
$V$ resonance in the $s$ channel. However since the $V$ bosons
are strongly coupled to longitudinal $W$'s, unless one imposes suitable cuts
in the $\cos\theta_W$ variable for the unpolarized 
differential cross section$^{6}$, the deviations will be clearly
visible only in the final $W_LW_L$ channel. 
 
To see the order of magnitude of the 
effect, we have considered the process
$e^+e^- \rightarrow W^+_{L} W^-_{L}{\bar \nu}\nu$, even if
it is perhaps unlikely that the experiments
will allow to reconstruct
such a distribution.  We have only applied a minor cut in the transverse
momentum of the outgoing $W$'s: $p_T\ge 10~GeV$.
 
The results for two options of $\sqrt{s}$ and $M_V$ are shown in
Table 1. The number of events are compared with
the corresponding SM predictions, obtained with $M_H=100~GeV$.
No branching ratio and selection efficiency has been
applied to these numbers. 
 
\def\dfl{\partial_{\mu}\mkern -18mu\raise 8pt\hbox{$\leftrightarrow$}}
\def\tvi{\vrule height 12pt depth 5pt width 0pt}
\def\tv{\tvi\vrule}
$$\vbox{\offinterlineskip\halign{\tv\quad#\quad\hfill&
\tv\quad#\quad\hfill&\tv\quad#\quad\hfill&\tv\quad#\quad\hfill&
\tv\quad#\quad\hfill&\tv\quad#\quad\hfill\tv
\cr  
\noalign{\hrule}
\hfill $\sqrt{s}$ &\hfill $M_{V}$& 
\hfill $(M_{WW})_{win.}$ &
\hfill $\#$ evts.&
\hfill$\#$ evts.&
\hfill $S/\sqrt B$\cr
\hfill $(TeV)$ &\hfill $(TeV)$ &\hfill $(TeV)$ 
&\hfill SM&\hfill BESS &\hfill \cr \noalign{\hrule}
\hfill 1.5&\hfill 1&\hfill 0.9~-~1.1&\hfill 7.7&\hfill 30.1&\hfill 11.5\cr
\hfill 2&\hfill 1.5&\hfill 1.3~-~1.6&\hfill 4.6&\hfill 15.3&\hfill 5.0\cr
\noalign{\hrule}}}$$
 
\baselineskip=10pt
\smallskip
\noindent
{\bf Table 1} - {\it  Fusion process $e^+e^- \rightarrow W^+_{L} 
W^-_{L}{\bar \nu}\nu$ for $L=80~fb^{-1}$ and $\sqrt s=1.5$, $2~TeV$.
$\Gamma_V$ is the width of the $V$ resonance; in the third column we give the
 window for the integration of $d\sigma (LL)/d M_{WW}$ on $d M_{WW}$.
 The last column shows
 the statistical significance.}
 
\smallskip
 
\baselineskip=14pt
 
 The two cases quoted in Table 1, corresponds to
$g/\gs=0.08$, $b=0.02$ (the first one) and $g/\gs=0.05$, $b=0.01$
(the second one).
 
The statistical significance looks good, but we must notice that it 
gets reduced when including 
the branching ratio and selection efficiency for the final
$W$'s (which is $\simeq 0.25$ for the decay in two jets). Therefore an even
larger luminosity is in general necessary to investigate the fusion channel$^6$.
 
Our conclusion is that, to test the SM against a possible 
electroweak strong sector, as described here through the BESS model, at these 
colliders, the annihilation channels are by far the most important.
 
Even if the mass of the $V$ resonance is
bigger than the c.m. energy of the collider, the process of $W$
pair production will allow for strong restrictions on the parameter 
space of the BESS model, especially so if the $W$ polarizations can be 
reconstructed from their decay distributions.
 
If no deviation from the SM prediction is found, already at $\sqrt{ s}=500~GeV$
and integrated luminosity $L=20~fb^{-1}$, 
the BESS model parameters $\gs$ and $b$ can be severely
restricted and we find significant improvement with respect to
LEP. With higher energy
colliders the parameter space contracts and at $b=0$
we get an upper bound on $g/\gs$ which for $M_V >\sqrt s$ is almost independent
on $M_V$.
 
\vskip 1truecm
\newcount \nref

\def\ref {\global \advance \nref by 1 \ifnum\nref<10 \item {$ \the\nref.~$}
\else \item{$\the\nref.~$} \fi}
\def\hb{\hfill\break}
\noindent
{\bf 4. References}
\vskip .5truecm
 
\ref
R. Casalbuoni, P. Chiappetta, S. De Curtis, D. Dominici, F. Feruglio
and R. Gatto, in "$e^+e^-$ Collisions at 500 $GeV$: the Physics Potential",
Proceedings of the Workshop, edited by P.M. Zerwas, DESY 92, 123B, 
August 1992, p. 513;\hb
D. Dominici, in "Physics and Experiments with Linear Colliders", 
Saariselk\"a, Finland, September 9-14, 1992, edited by R. Orava, P. Eerola
and M. Nordberg, World Scientific, p. 509.      
\ref 
   R. Casalbuoni, P. Chiappetta, A. Deandrea, S. De Curtis, D. Dominici, 
     and R. Gatto, UGVA-DPT 1993/02-805, February 1993.
\ref
    R. Casalbuoni, S. De Curtis, D. Dominici and R. Gatto, {\it Phys. Lett.}
    {\bf B155} (1985) 95;
    {\it Nucl. Phys.} {\bf B282} (1987) 235.
\ref
    C. De Clercq, to appear on the proceedings of the XXVIII Rencontres de
    Moriond on Electroweak Interactions, Les Arcs, March 1993;\hb
    V. Innocente, {\it ibidem};\hb
    R. Tenchini, {\it ibidem};\hb
    G. Altarelli, talk given at  "FILEP - Incontro sulla Fisica a LEP"
    Firenze 1-2 Aprile 1993.
\ref   
K. Fujii, KEK preprint 92-31, to appear in the Proceedings of
the 2nd KEK Topical
Conference on $e^+e^-$ Collision Physics,
KEK, Tsukuba, Japan, November 26-29 1991. 
 
\ref   
Y. Kurihara, Contribution to these proceedings;\hb
T. Han, Contribution to these proceedings.
 
\vfill\eject
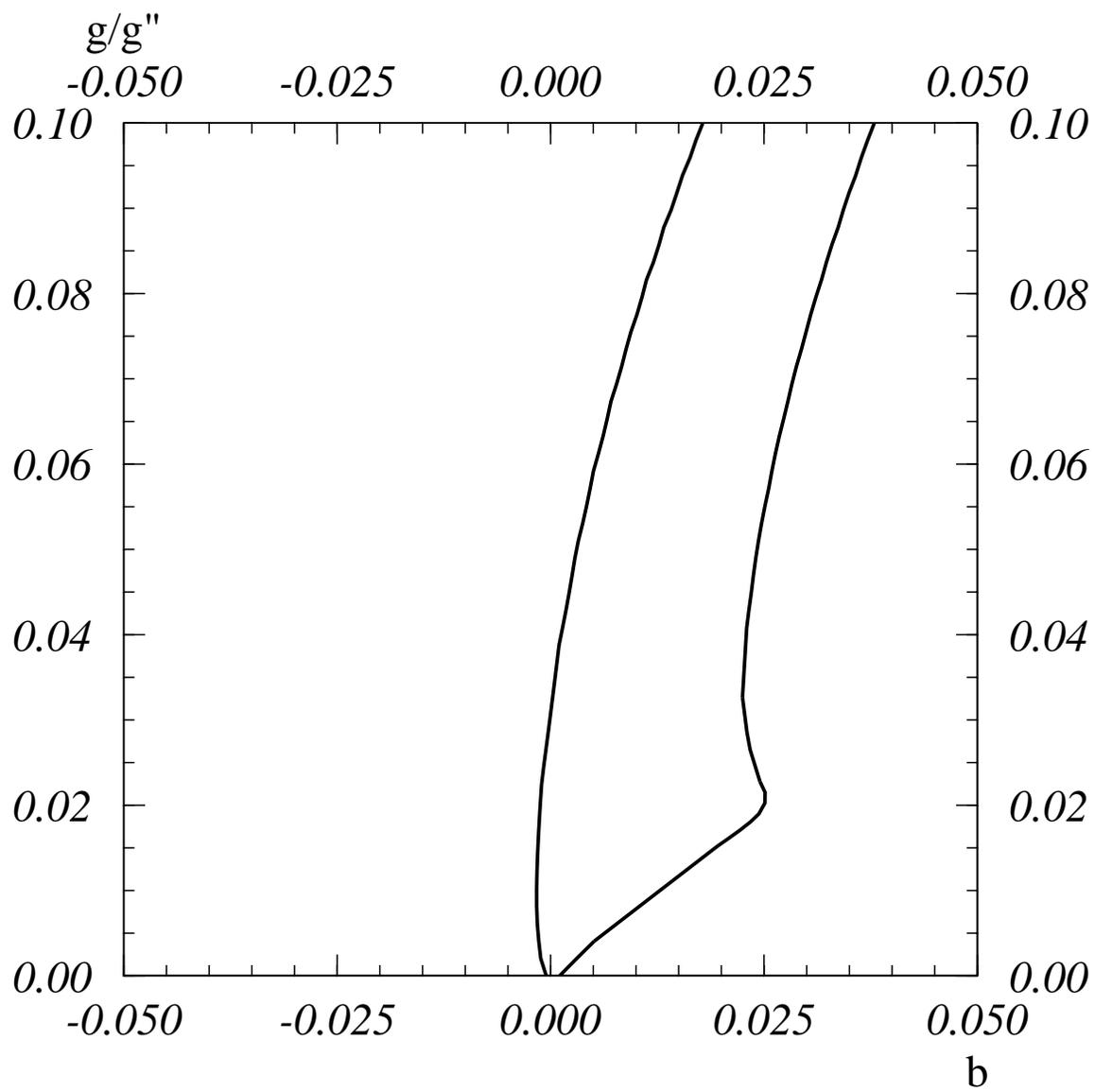
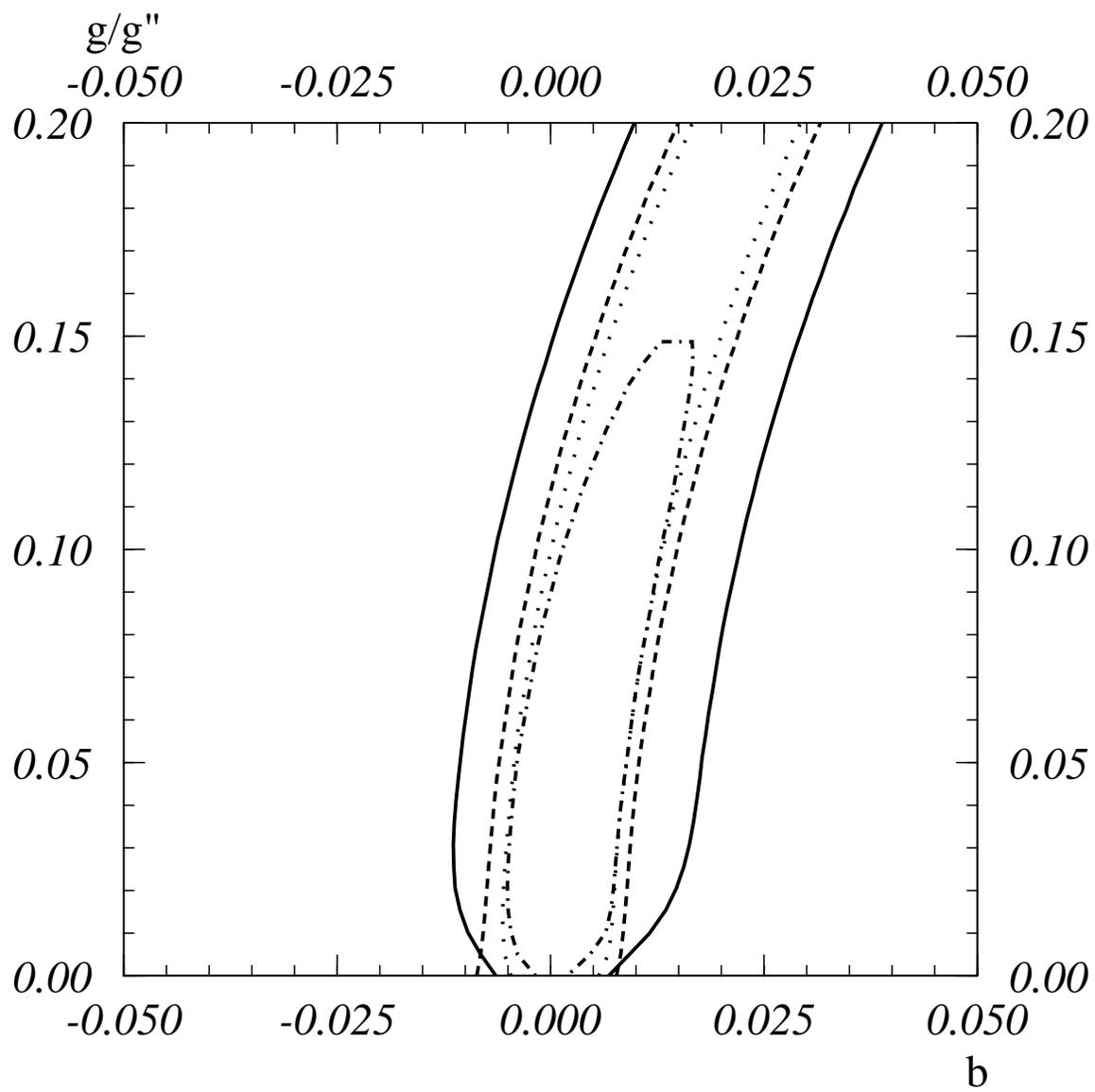
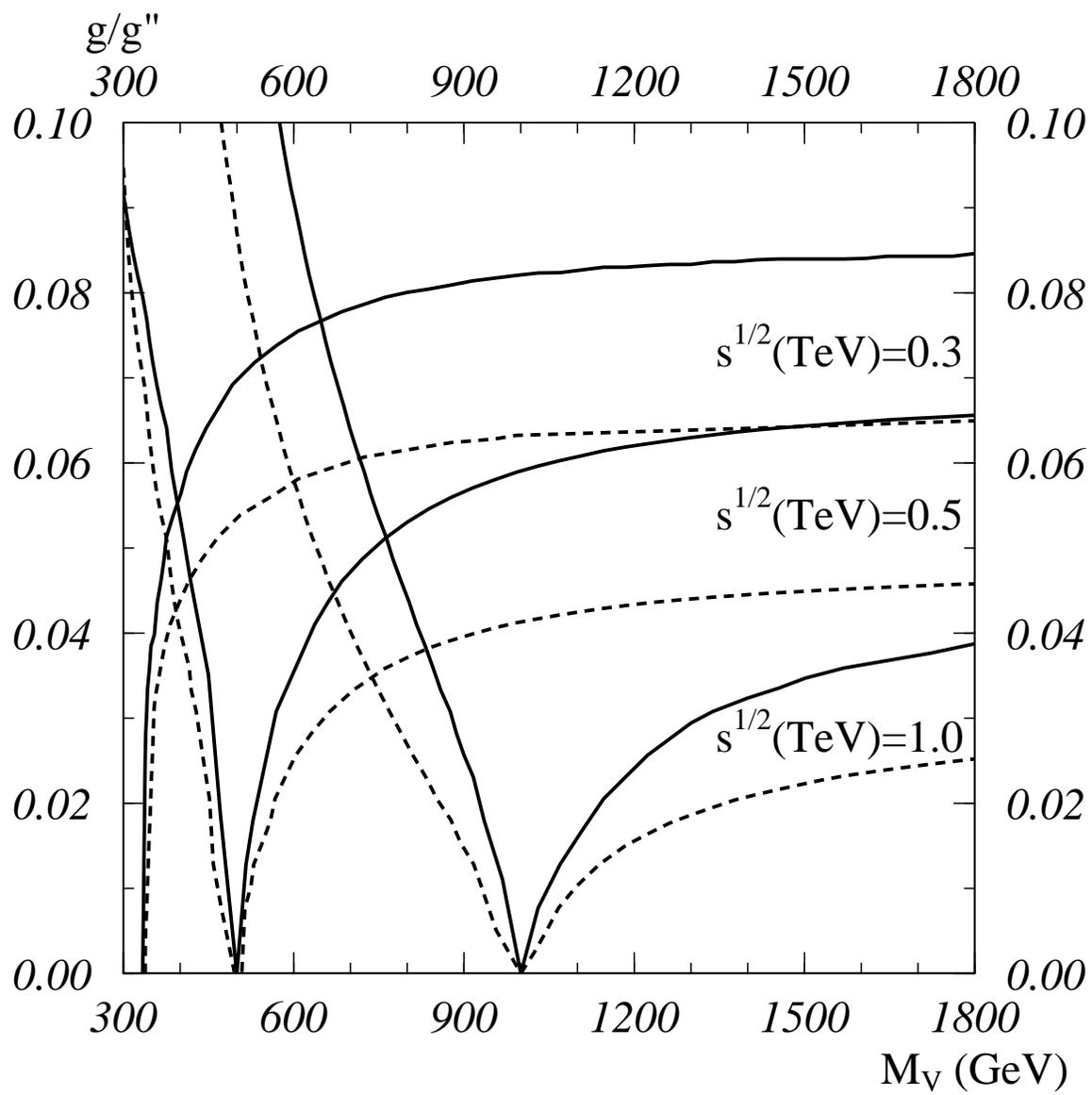
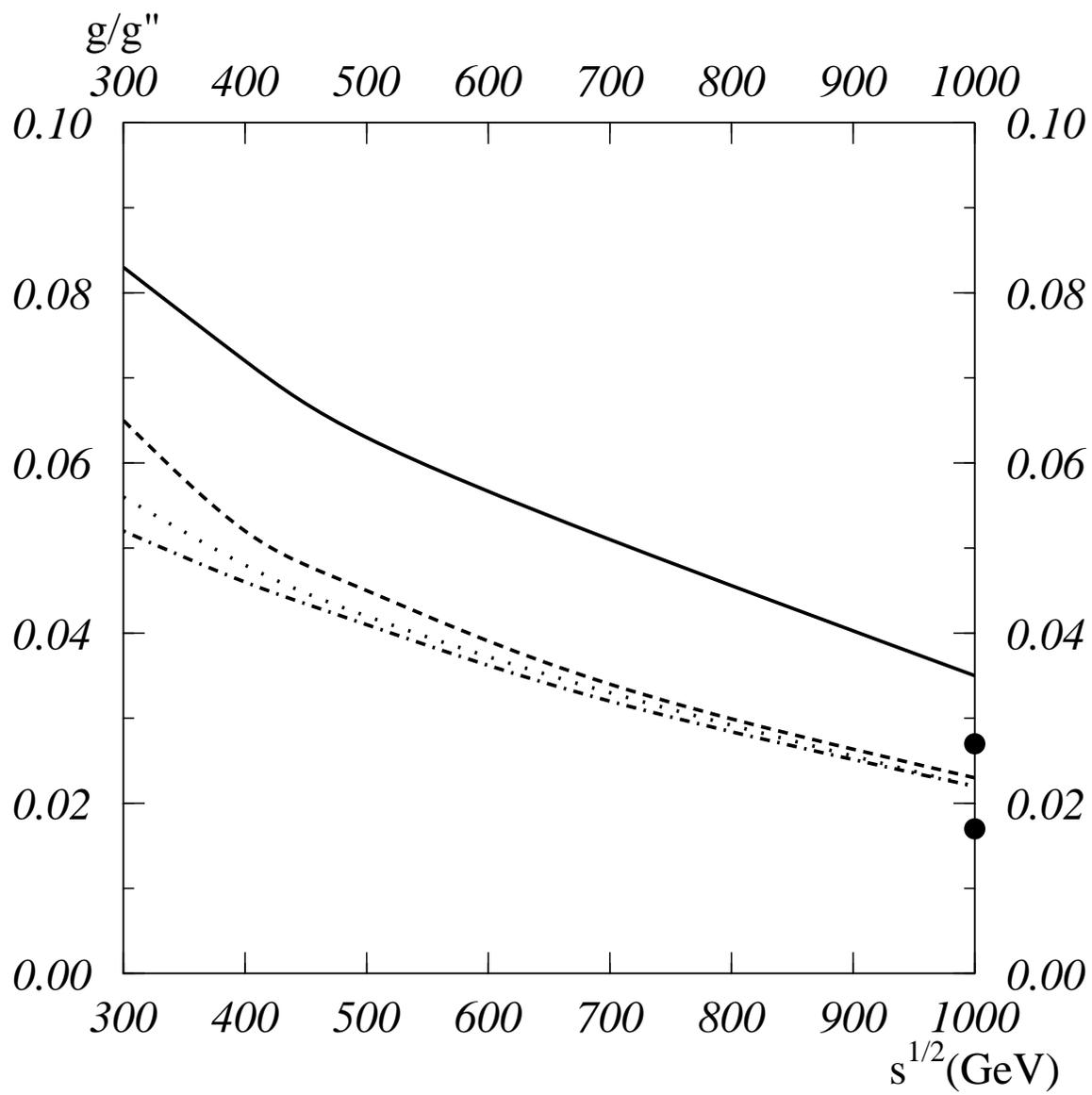
\bye